\documentclass[10ps]{article}

\pdfoutput=1

\usepackage{isabelle,isabellesym}
\usepackage{graphicx}
\usepackage{amsmath} 
\usepackage{amsfonts}
\usepackage{amssymb}
\usepackage{amscd}
\usepackage{amsthm}
\usepackage{latexsym}
\usepackage{mathrsfs}
\usepackage{stmaryrd}
\usepackage{mathpartir} % inferrules
\usepackage{hyperref}
\usepackage{color}
\usepackage{tikz}
\usepackage{tikz-cd}
\usepackage{fullpage}
\usepackage[colorinlistoftodos]{todonotes}
\usepackage{hyperref}

\newcommand{\IF}[3]{\mathbf{if}\ #1\ \mathbf{then}\ #2\ \mathbf{else}\ #3}
\newcommand{\WHILE}[2]{\mathbf{while}\ #1\ \mathbf{do}\ #2}
\newcommand{\WHILEI}[3]{\mathbf{while}\ #1\ \mathbf{inv}\ #2\ \mathbf{do}\ #3}
\newcommand{\sskip}{\mathit{skip}}

\newcommand{\KAT}{\mathsf{KAT}}
\newcommand{\rKAT}{\mathsf{rKAT}}

\newcommand{\dL}{\mathsf{d}\mathcal{L}}
\newcommand{\dH}{\mathsf{d}\mathcal{H}}
\newcommand{\dR}{\mathsf{d}\mathcal{R}}
\newcommand{\flow}{\varphi}

\newcommand{\Pow}{\mathcal{P}}

\newcommand{\reals}{\mathbb{R}}
\newcommand{\bools}{\mathbb{B}}

\newcommand{\Sols}{\mathop{\mathsf{Sols}}}
\newcommand{\sta}{\mathsf{Sta}}

\definecolor{scolor}{rgb}{1,0.5,0.5}
\definecolor{jcolor}{cmyk}{1,0,1,0}
\definecolor{gcolor}{cmyk}{1,0,0,0}

\newcommand{\lput}{\textit{\textsf{put}}}
\newcommand{\lget}{\textit{\textsf{get}}}

\newcommand{\lto}{\Longrightarrow}

\newcommand{\lindep}{\mathop{\,\bowtie\,}}

\newcommand{\unrest}{\mathop{\sharp}}
\newcommand{\seq}{\mathop{\,;\,}}
\newcommand{\subapp}{\mathop{\,\dagger\,}}

\newcommand{\isactrlU}{\textbf{\textsf{U}}}

\newtheorem{proposition}{Proposition}[section]
\newtheorem{lemma}{Lemma}[section]

\theoremstyle{definition}

\newtheorem{example}{Example}[section]

\isabellestyle{it}

\begin{document}

\title{Differential Hoare Logics and Refinement Calculi\\ for Hybrid Systems with Isabelle/HOL}

%\titlerunning{Hoare Logics and Refinement Calculi for Hybrid Systems}

\author{Simon Foster\\ University of York\\
United Kingdom \and Jonathan Juli\'an Huerta y Munive\\ University of Sheffield\\
United Kingdom \and Georg Struth\\ University of Sheffield\\
United Kingdom} 
%\authorrunning{Foster, Huerta y Munive and Struth}

%\institute{University of York, UK \and University of Sheffield, UK}

\maketitle

\begin{abstract} 
  We present simple new Hoare logics and refinement calculi for hybrid
  systems in the style of differential dynamic logic. (Refinement)
  Kleene algebra with tests is used for reasoning about the program
  structure and generating verification conditions at this level.
  Lenses capture hybrid program stores in a generic algebraic way. The
  approach has been formalised with the Isabelle/HOL proof
  assistant. A number of examples explains the workflow with the
  resulting verification components.

\vspace{\baselineskip}

\end{abstract}

%%%%%%%%%%%%%%%%%%%%%%%%%%%%%%%%%%%%%%%%%%%%%%%%%%

\section{Introduction}\label{sec:introduction}

Differential dynamic logic ($\dL$) is a prominent deductive method for
verifying hybrid systems~\cite{Platzer18}. It extends dynamic logic
with domain-specific inference rules for reasoning about the discrete
control and continuous dynamics that characterise such
systems. Continuous evolutions are modelled by $\dL$'s evolution
commands within a hybrid program syntax. These declare a vector field
and a guard, which is meant to hold along the evolution.  Reasoning
with evolution commands in $\dL$ requires either explicit solutions to
differential equations represented by the vector field, or invariant
sets~\cite{Teschl12} that describe these evolutions implicitly. Verification
components inspired by $\dL$ have already been formalised in the
Isabelle proof assistant~\cite{MuniveS19}. Yet the shallow embedding
used has shifted the focus from the original proof-theoretic approach
to a semantic one, and ultimately to predicate transformer algebras
supporting a quite different workflow.

Dynamic logics and predicate transformers are powerful tools. They
support reasoning about program equivalences and transformations
far beyond what standard program verification requires~\cite{BackW98}. For
the latter, much simpler Hoare logics generate precisely the
verification conditions needed.  Asking about the feasibility of a
\emph{differential Hoare logic} ($\dH$) is therefore natural and
interesting.  As Hoare logic is strongly related to Morgan's
refinement calculus~\cite{Morgan94}, it is equally reasonable to
ask whether and how a Morgan-style \emph{differential refinement calculus}
($\dR$) might allow constructing hybrid programs from specifications.

A prima facie answer to these questions seems positive: after all, the
laws of Morgan's refinement calculus can be proved using the rules of
Hoare logic, which in turn are derivable within dynamic logic. But the
formalisms envisaged might not be expressive enough for hybrid program
verification or less suitable than $\dL$ in practice. Conceptually it
is also not obvious what exactly it would take to extend a standard
Hoare logic or refinement calculus to hybrid programs.

Our main contribution consists in evidence that $\dH$ and $\dR$ are as
feasible and applicable for verifying simple hybrid programs as $\dL$,
and that developing these novel methods requires simply adding a
single Hoare-style axiom and a single refinement rule for evolution
commands to the standard formalisms.

This conceptual simplicity is reflected in the Isabelle verification
components for $\dH$ and $\dR$. These reuse components for
(refinement) Kleene algebra with
tests~\cite{Kozen97,ArmstrongGS16,afp:vericomp} $(\mathsf{(r)KAT}$)
for the propositional Hoare logic and refinement calculi---ignoring
assignment and evolution commands. The axioms and laws for these two
basic commands are derived in a concrete state transformer semantics
for hybrid programs~\cite{afp:hybrid} over a generic hybrid
store model based on lenses~\cite{FosterZW16}, reusing other Isabelle
components~\cite{afp:hybrid,Foster18c-Optics,Foster19a-IsabelleUTP}.
Data-level verification conditions are discharged using Isabelle's
impressive components for ordinary differential
equations~\cite{ImmlerH12a}.

This simple modular development evidences the benefits of algebraic
reasoning and shallow embeddings with proof assistants. Our
verification components merely require formalising a state transformer
semantics for $\KAT$ and $\rKAT$ along the lines of~\cite{MuniveS19}
and concrete store semantics for hybrid
programs. Lenses~\cite{FosterZW16} give us the flexibility to switch
seamlessly between stores based on real vector spaces or executable
Euclidean spaces. Beyond that it suffices to derive a few algebraic
laws for invariants and the Hoare-axioms and refinement laws for
evolution commands in the concrete semantics. Program verification is
then performed at the concrete level, but this remains hidden, as
tactics generate data-level verification conditions automatically and
we have programmed boiler-plate syntax for programs and correctness
specifications.

Our Isabelle components support the  workflows of $\dL$ in $\dH$
and $\dR$. We may reason explicitly with solutions to differential
equations and implicitly with invariant sets. We have formalised a third
method in which solutions, that is flows, are declared ab initio in
correctness specifications and need not be certified.

Our program construction and verification components have so far been
evaluated on a small set of simple examples. We present some of them
to explain the work flows supported by $\dH$ and $\dR$. With Isabelle
tactics for automated verification condition generation in place, we
notice little difference relative to our predicate transformer
components~\cite{MuniveS19}.  The entire Isabelle formalisation is
available
online\footnote{\url{https://github.com/yonoteam/HybridKATpaper}}; our
repository contains instructions for using it and a postscript proof
document for reading the Isabelle code without installing the tool.

%%%%%%%%%%%%%%%%%%%%%%%%%%%%%%%%%%%%%%%%%%%%%%%%%%

\section{Kleene Algebra with Tests}\label{sec:kat} 

A \emph{Kleene algebra with tests}~\cite{Kozen97} ($\KAT$) is a
structure $(K,B,+,\cdot,0,1,^\ast,\neg)$ where $(B,+,\cdot,0,1,\neg)$
is a boolean algebra with join $+$, meet $\cdot$, complementation
$\neg$, least element $0$ and greatest element $1$, $B\subseteq K$,
and $(K,+,\cdot,0,1,^\ast)$ is a Kleene algebra---a semiring with
idempotent addition equipped with a star operation that satisfies the
axioms $1+\alpha\cdot\alpha^\ast \le \alpha^\ast$ and
$\gamma+\alpha\cdot \beta\le \beta\rightarrow \alpha^\ast \cdot
\gamma\le \beta$,
as well as their opposities, with multiplication swapped.  The
ordering on $K$ is defined by
$\alpha\le \beta\leftrightarrow \alpha+\beta=\beta$, as idempotent
semirings are semilattices. We often write $\alpha\beta$ instead of
$\alpha\cdot\beta$, and use $p,q,r,\dots$ for elements of $B$.

Elements of $K$ represent programs; those of $B$ tests, assertions or
propositions.  The operation $\cdot$ models the sequential composition
of programs, $+$ their nondeterministic choice, $(-)^\ast$ their
finite unbounded iteration. Program $0$ aborts and $1$ skips.  Tests
are embedded implicitly into programs. They are meant to hold in some
states of a program and fail in others; $p\alpha$ ($\alpha p$)
restricts the execution of program $\alpha$ in its input (output) to
those states where test $p$ holds. The ordering $\le$ is the opposite
of the refinement ordering on programs (see Section~\ref{sec:refine}).

Binary relations of type $\Pow\, (S\times S)$ form
$\KAT$s~\cite{Kozen97} when $\cdot$ is interpreted as
relational composition, $+$ as relational union, $(-)^\ast$
as reflexive-transitive closure and the elements of $B$ as
subidentities---relations below the relational unit. This
grounds $\KAT$ within standard relational imperative program
semantics. However, we prefer the isomorphic representation known
as \emph{state transformers} of type $S\to \Pow\, S$.
Composition $\cdot$ is then interpreted as Kleisli
composition
\begin{equation*} 
(f\circ_K g)\, x = \bigcup\{g\, y\mid y \in f\ x \}, 
\end{equation*} 
$0$ as $\lambda x.\ \emptyset$ and $1$ as $\eta_S = \{-\}$.  Stars
$f^{\ast}\, s  = \bigcup_{i\in\mathbb{N}} f^i\, s$ are defined with
  respect to Kleisli composition using $f^{0} = \eta_S$ and
  $f^{n+1} = f \circ_K f^{n}$. The boolean algebra of tests has
  carrier set $B_S=\{f:S\to \Pow\, S \mid f\le \eta_S\}$, where the
  order on functions has been extended pointwise,  and complementation
  is given by
  \begin{equation*} 
    \overline{f}\, x =
  \begin{cases}
    \eta_S\, x, & \text{ if } f\, x = \emptyset,\\
\emptyset, & \text{ otherwise}.
  \end{cases}
\end{equation*}
We freely identify predicates, sets and state 
transformers below $\eta_S$, which are isomorphic:
$P\cong \{s\mid P\, s\}\cong \lambda s.\ \{x\mid x=s \land P\, s\}$.

\begin{proposition}\label{P:kleisli-ka}
$\sta\, S = ((\Pow\, S)^S,B_S,\cup,\circ_K,\lambda x.\
  \emptyset, \eta_S,(-)^{\ast},\overline{(-)})$
  forms a $\KAT$, the \emph{full state transformer $\KAT$} over the
  set $S$.
\end{proposition}
A \emph{state transformer $\KAT$} over $S$ is any subalgebra of
$\sta\, S$. 

We have already formalised $\KAT$ via type classes in
Isabelle~\cite{afp:kat}.  As these allow only one type parameter, we
use an alternative approach that expands a Kleene algebra $K$ by an
\emph{antitest} function $n:K\to K$ from which a \emph{test} function
$t:K\to K$ is defined as $t=n^2$. Then
$K_t = \{\alpha \mid t\, \alpha = \alpha\}$ forms a boolean algebra in
which $n$ acts as test complementation. It can be used in place of
$B$.  The formalised state transformer model of $\KAT$ is a
contribution to this article.

%%%%%%%%%%%%%%%%%%%%%%%%%%%%%%%%%%%%%%%%%%%%%%%%%%

\section{Propositional Hoare Logic and Invariants}\label{sec:hl-invariants}

$\KAT$ provides a simple algebraic semantics for while
programs with
\begin{align*}
  \IF{p}{\alpha}{\beta} = p\cdot \alpha + \neg p \cdot
  \beta\qquad\text{ and }\qquad
\WHILE{p}{\alpha} = (p\cdot \alpha)^\ast \cdot \neg p.
\end{align*}
It captures validity of Hoare triples in
a partial correctness semantics as
\begin{equation*}
  \{p\}\, \alpha\, \{q\} \leftrightarrow p\alpha\neg q = 0,
\end{equation*}
or equivalently by $p\alpha\le \alpha q$ or $p\alpha = p\alpha q$. It
also allows deriving the rules of \emph{propositional Hoare
  logic}~\cite{Kozen00}---disregarding assignments---which are useful for
verification condition generation:
\begin{align}
  &\{p\}\, \sskip\, \{p\}, \label{eq:h-skip}\tag{h-skip}\\
  p\le p' \land \{p'\}\, \alpha\, \{q'\} \land q'\le q\ \rightarrow\ &
                                                                       \{p\}\,
                                                                       \alpha\,
                                                                       \{q\},\label{eq:h-cons}\tag{h-cons}\\
  \{p\}\, \alpha\, \{r\} \land \{r\}\, \beta\, \{q\}\ \rightarrow\
  &\{p\}\, \alpha\beta\, \{q\},\label{eq:h-seq}\tag{h-seq}\\
  \{tp\}\, \alpha\, \{q\}\land \{\neg tp\}\, \beta\, \{q\}\
  \rightarrow\ & \{p\}\, \IF{t}{\alpha}{\beta}\, \{q\},\label{eq:h-cond}\tag{h-cond}\\
  \{tp\}\, \alpha\, \{p\}\ \rightarrow\ & \{p\}\, \WHILE{t}{\alpha}\, \{\neg tp\}.\label{eq:h-while}\tag{h-while}
\end{align}

Rules for commands with invariant assertions $\alpha\ \mathbf{inv}\ i$
are derivable, too (operationally,
$\alpha\, \mathbf{inv}\, i = \alpha$).  An \emph{invariant} for
$\alpha\in K$ is a test $i\in B$ satisfying $\{i\}\, \alpha\, \{i\}$.
Then, with $\mathbf{loop}\, \alpha$ as syntactic sugar for
$\alpha^\ast$, we obtain
\begin{align}
  p\le i \land \{i\}\, \alpha\, \{i\}\land i\le q\ \rightarrow\
  &\{p\}\, \alpha\, \{q\},\label{eq:h-inv}\tag{h-inv}\\
  \{i\}\, \alpha\, \{i\} \land \{j\}\, \alpha\, \{j\}\rightarrow\
  &\{i j\}\, \alpha\, \{i j\},\label{eq:h-inv-mult}\tag{h-inv-mult}\\
  \{i\}\, \alpha\, \{i\} \land \{j\}\, \alpha\, \{j\}\rightarrow\
  &\{i+ j\}\, \alpha\, \{i+ j\},\label{eq:h-inv-plus}\tag{h-inv-plus}\\
  p \le i \wedge \{it\}\, \alpha\, \{i\} \wedge \neg t i\le q\
  \rightarrow \ & \{p\}\, \WHILEI{t}{i}{\alpha}\,  \{q\},\label{eq:h-while-inv}\tag{h-while-inv}\\
   p\le i \land \{i\}\, \alpha\, \{i\}\land i\le q\ \rightarrow\ &
                                                                   \{p\}\, \mathbf{loop}\, \alpha\,
    \mathbf{inv}\, i\, \{q\}. \label{eq:h-loop-inv}\tag{h-loop-inv}
\end{align}
We use (\ref{eq:h-inv}) for invariants for continuous
evolutions of hybrid systems in Section~\ref{sec:hoare-inv}-\ref{sec:from-flows}. The rules (\ref{eq:h-inv-mult}) and (\ref{eq:h-inv-plus}) are part of a procedure, described in Section~\ref{sec:hoare-inv}. Rule (\ref{eq:h-while-inv})
is standard for  invariants for while loops;
(\ref{eq:h-loop-inv}) is specific to loops of hybrid programs (see
Section~\ref{sec:sta-hybrid}).

The rules for propositional Hoare logic in Isabelle have been derived
for $\KAT$ in~\cite{afp:kat,afp:vericomp}. The rules for invariants
have been developed specifically for this article.

%%%%%%%%%%%%%%%%%%%%%%%%%%%%%%%%%%%%%%%%%%%%%%%%%%

\section{State Transformer Semantics for Hybrid
  Programs}\label{sec:sta-hybrid}

Hybrid programs of differential dynamic logic ($\dL$)~\cite{Platzer18}
are defined by the syntax
\begin{equation*}
\mathcal{C}\ ::= \ x:=e \mid x' = f \, \&\, G \mid ?P\mid \mathcal{C};\mathcal{C}\mid \mathcal{C}+\mathcal{C}\mid \mathcal{C}^*
\end{equation*}
that adds \emph{evolution commands} $x' = f \, \&\, G$ to the language
of $\KAT$---function $?(-)$ embeds tests explicitly into programs.
Evolution commands introduce a time independent vector field $f$ for
an autonomous system of ordinary differential equations
(ODEs)~\cite{Teschl12} together with a guard $G$, a predicate
modelling boundary conditions or similar restrictions on temporal
evolutions. % Guards are also known as \emph{evolution domain
  % restrictions}~\cite{DoyenFPP18}.

Formally, we fix a state space $S$ of the hybrid program,
for example $S\subseteq \reals^n$ and $n\in\mathbb{N}$. We
model continuous variables algebraically using
lenses~\cite{FosterZW16} to support different state space
models generically. A lens, $x : A \lto S$, is a tuple
$x = (A, S, \lget, \lput)$ with variable type $A$ and state
space $S$. The functions $\lget_x : S \to A$ and
$\lput_x : S \to A \to S$ query and update the value of $x$
in a particular state. They are linked by three intuitive
algebraic laws~\cite{FosterZW16}:
\begin{equation*}
\lget~(\lput~s~v) = v, \qquad \lput~(\lput~s~v')~v = \lput~s~v, \qquad
\lput~s~(\lget~s) = s,
\end{equation*}
where $s \in S$ and $v, v' \in A$.

Lenses $x$ and $y$ can also be checked for independence
using the predicate $x \lindep y$, which we use to
distinguish variables. Each program variable is a lens
$x : \reals \lto S$. State spaces of the form
$S \subseteq \reals^n$ thus have $n$ independent lenses
$x_1 \cdots x_n$ corresponding to projections from
$\reals^n$. Yet more general state spaces such as vector spaces or
infinite Euclidean spaces can be supported as well.

Systems of equations are modelled using vector fields: functions of
type $S\to S$ on some open set $S$. Geometrically, vector field $f$
assigns a vector to any point of the state space $S$. A solution to
the \emph{initial value problem} (IVP) for the pair $(f,s)$ and
initial value $(0,s)\in T\times S$, where $T$ is an open interval in
$\reals$ containing $0$, is then a function $X:T\to S$ that satisfies
$X'\, t = f\, (X\, t)$---an autonomous system of ODEs in vector
form---and $X\, 0 = s$. Solution $X$ is thus a curve in $S$ through
$s$, parametrised in $T$ and tangential to $f$ at any point in $S$; it
is called a \emph{trajectory} or \emph{integral curve} of $f$ at $s$
whenever it is uniquely defined~\cite{Teschl12}.

For IVP $(f,s)$ with continuous vector field $f:S\to S$ and initial
state $s\in S$ we define the set of solutions on
$T$ as
\begin{equation*}
\Sols f\, T\, s = \left\{X \mid \forall t\in T.\  X'\, t = f\, (X\, t)\land X\, 0 = s\right\}.
\end{equation*}
Each solution $X$ is then continuously differentiable and thus
$f\circ X$ integrable in $T$.  For $X\in \Sols\, f\, T\, s$ and
$G:S\to\bools$, we further define the $G$-\emph{guarded orbit} of $X$
along $T$ in $s$~\cite{MuniveS19} with the help of the state transformer
$\gamma^X_G:S\to \Pow\, S$ as 
\begin{equation*}
\gamma^X_{G}\, s= \left\{X\, t\mid t\in T\land \forall \tau\in
{\downarrow}t.\ G\, (X\, \tau)\right\},
\end{equation*}
where ${\downarrow}t = \left\{t'\in T\mid t'\le t\right\}$, and the
$G$-\emph{guarded orbital} of $f$ along $T$ in $s$~\cite{MuniveS19}
via the state transformer $\gamma^f_G:S\to \Pow\, S$ as
\begin{equation*}
  \gamma^f_G\ s = \bigcup\left\{\gamma^X_G\, s\mid X\in \Sols\, f\, T\, s\right\}.
\end{equation*}
In applications, ${\downarrow}t$ is usually an interval
$[0,t]\subseteq T$.  Expanding definitions,
\begin{equation*}
\gamma^f_G\, s = \left\{X\, t \mid X\in \Sols\, f\, T\, s \land t\in T
\land \forall \tau\in{\downarrow}t.\ G\, (X\, \tau)\right\}.
\end{equation*}
If $\top$ denotes the predicate that holds of all states in $S$ (or
the set $S$ itself), we write $\gamma^f$ instead of
$\gamma^f_\top$. We define the semantics of the evolution command
$x'= f\, \&\, G$~\cite{MuniveS19} for any continuous $f:S\to S$ and
$G:S\to \bools$ as
\begin{equation}
{\left(x'= f\, \&\, G\right)} = \gamma^f_G.\label{eq:st-evl}\tag{st-evl}
\end{equation}

Defining the state transformer semantics of assignments is
standard~\cite{MuniveS19}, though we generalise using lenses. First,
we use lenses to define state updates:
$$\sigma(x\mapsto e) = \lambda s.\ \lput_x ~ (\sigma~s) ~ (e~s)$$
for $x : A \lto S$, $e : S \to A$, and $\sigma : S \to S$.
Intuitively, this updates the value of variable $x$ in the state
update $\sigma : S \to S$ to have the value given by $e$. Here, the
function $e$ models an ``expression'' that is evaluated in
state $s$. For example, if $x$ and $y$ are
variables, then the expression $x / (2 + y)$ is modelled by
$\lambda s. \ \lget_x~s ~/~ (2 + \lget_y~s)$. We can also update $n$
variables simultaneously:
$$[x_1 \mapsto e_1, x_2 \mapsto e_2, \cdots, x_n \mapsto e_n] = \textit{id}(x_1 \mapsto e_1)(x_2 \mapsto e_2)\cdots(x_n \mapsto e_n),$$ 
where $\textit{id}$ is the identity function. State updates
commute, when assigning to independent lenses, and cancel
one another out, when made to the same lens, in a natural
way. We also define a substitution operator for state
updates, $\sigma \dagger e = e \circ \sigma$, that simply
composes expression $e$ with $\sigma$. We can then define a
semantic analog of the substitution operator,
$e[f/x] = [x \mapsto f] \subapp e$ that satisfies the
expected laws~\cite{FosterZW16}. Finally, we define a
generalised assignment operator:

\begin{equation}
  \langle \sigma \rangle = \lambda s.\ \{ \sigma(s) \}.\tag{st-assgn}
\end{equation}
This applies $\sigma : S \to S$ as an assignment. With our state
update function, singleton assignment is a special case:
$(x := e) =\langle x \mapsto e \rangle$, and we can also assign
several variables simultaneously. These foundations allow us to derive
standard laws for assignments algebraically, as for instance in
schematic $\KAT$~\cite{AngusK01}:
\begin{align*}
  x := x &~=~ \sskip, \\
  x := e \seq x := f &~=~ x := f[e/x], \\
  x := e \seq y := f &~=~ y := f \seq x := e, \qquad\qquad
                       \textnormal{if~} x \lindep y, x \unrest f, y
                       \unrest e, \\
  x := e \seq \textbf{if}~ t ~ \textbf{then}~ \alpha ~ \textbf{else}~ \beta 
                          &~=~ \textbf{if}~ t[e/x]  ~\textbf{then}~ x := e \seq \alpha  ~\textbf{else}~ x := e \seq \beta.
\end{align*}

Here, $x \unrest e$ means that the semantic expression $e$
does not depend in its valuation on lens
$x$~\cite{FosterZW16}. An assignment of $x$ to itself is
simply $\sskip$. Two assignments to $x$ result in a single
assignment, with a semantic substitution
applied. Assignments to independent variables $x$ and $y$
commute provided that neither assigned expression depend on
the corresponding variable.  Assignment can be distributed
through conditionals by a substitution to the
condition. Such laws can be applied recursively for symbolic
evaluation of deterministic programs.

% First we define a state update function
% $f_a:V\to (S \to E) \to S\to S$ as
% \begin{equation*}
% f_a\, x\, e\, s = s[x\mapsto e\, s],
% \end{equation*}
% where $f[a\mapsto b]$ updates $f:A\to B$ by associating $a\in A$ with
% $b$ and every $y\neq a$ with $f\, y$.  The ``expression''
% ${e:S\to \reals}$ is thus evaluated in state $s$ to $e\, s$.  Then we
% lift $f_a\, x\, e:S\to S$ to state transformer
% $x:= e:S \to \Pow\, S$ using $\eta_S$:
% \begin{equation}
%   (x:= e) = \lambda s.\ \{f_a\, x\, e\, s\}.\label{eq:st-assgn}\tag{st-assgn}
% \end{equation}

Lenses support various store models, including records and
functions~\cite{FosterZW16}. We provide models for
vector spaces, executable and infinite Euclidean spaces:
\begin{align*}
  \textit{vec-lens}^n_k &=  (\reals, \reals^n, \lambda s.\, \textit{vec-nth}~s~k, \lambda s~v.\, \textit{vec-upd}~k~v~s), & \text{if~} k < n, \\
  \textit{eucl-lens}^n_k & = (\reals, V, \lambda s.\, \textit{eucl-nth}~s~k, \lambda s~v.\, \textit{eucl-upd}~k~v~s), & \text{if~} k < n, \\
%  \textit{eucl-lens}^{(m,n)}_k & = (\reals, M^{(m,n)}, \lambda s.\, \textit{eucl-nth}~s~k, \lambda s~v.\, \textit{eucl-upd}~k~v~s), & \text{if~} k < mn, \\
  \textit{fun-lens}^{(A,B)}_i &= (B, A \to B, (\lambda f. f i), (\lambda f~v. \, f(i := v))).
\end{align*}

\noindent The vector lens selects the $k$th element of an $n$
dimension vector using $\textit{vec-nth}$ and $\textit{vec-upd}$ from
the HOL Analysis library~\cite{HolzlIH13}, which provides an indexed
type for the space $\reals^n$. The Euclidean lens uses executable Euclidean
spaces~\cite{ImmlerT19} that provide a list representation of the vectors in the 
$n$-dimensional $V$ via an ordered basis and an inner product. The
function lens selects range elements of a function associated with a
domain element $i \in A$. It can be used in particular with infinite
Euclidean spaces, $\mathbb{N} \to \reals$. All three satisfy the lens
axioms above.

The development in this section has been formalised with
Isabelle~\cite{afp:hybrid,Foster18c-Optics,Foster19a-IsabelleUTP},
both for a state transformer and a relational semantics. An instance
of the latter for particular vector fields with unique solutions forms
the standard semantics of $\dL$. By the direct connection to orbits or
orbitals, the state transformer semantics is arguably conceptually
simpler and more elegant.

%%%%%%%%%%%%%%%%%%%%%%%%%%%%%%%%%%%%%%%%%%%%%%%%%%

\section{Differential Hoare Logic for Flows}\label{sec:hoare-flow}

In the standard semantics of Hoare triples, the Kleisli composition in
the left hand side of $p\alpha\le \alpha q$ ensures that $p$ holds
before executing $\alpha$. The left hand side guarantees that $q$ holds
after its execution. With evolution commands, and consistently with
$\dL$, the $q$ holds at every point in the orbit of a solution for $f$.

The assignment axiom of Hoare logic needs no explanation. Our concrete
semantics allows us to derive it:

\begin{equation}
\left\{P[e/x]\right\}\,  x:=e\, \{P\}. \label{eq:h-assgn}\tag{h-assgn}
\end{equation}

% Optimised assignment laws, in case it's needed.
% \begin{equation}
% \left\{x = e[x_0/x] \wedge p[x_0/x]\right\}, S \, \left\{q\right\} \Longrightarrow \left\{p \right\}\, x := e \seq S \,\left\{ q \right\}
% \end{equation}

Hence, all we need to add to Hoare logic is a rule for evolution
commands.  We restrict our attention to Lipschitz-continuous vector
fields for which unique solutions to IVPs are guaranteed by
Picard-Lindel\"of's theorem~\cite{Teschl12}.  These are \emph{(local)\
  flows} $\flow:T\to S\to S$ and $X=\flow_s=\lambda t.\ \flow\, t\, s$
is the trajectory at $s$. Guarded orbitals $\gamma^f_G$ then
specialise to \emph{guarded orbits}
\begin{equation*}
  \gamma^f_{G,U} = \left\{\flow_s\, t\mid t\in U\land \forall\tau \in
  {\downarrow}t.\ G\, (\flow_s\, t)\right\},
\end{equation*}
where $U\subseteq T$ is a time domain of interest, typically an
interval $[0,t]$ for some $t\in T$~\cite{MuniveS19}.  Accordingly,
(\ref{eq:st-evl}) becomes
\begin{equation}
  \left(x' = f\, \&\, G\right)= \gamma^f_{G,U}.\label{eq:st-evl-flow}\tag{st-evl-flow}
\end{equation}
The following Hoare-style rule for evolution commands is then
derivable.
\begin{lemma}\label{P:h-evl-lemma}
  Let $f:S\to S$ be a Lipschitz continuous vector field on
  $S\subseteq \reals^n$ and $\flow:T\to S\to S$ its local flow with
  $0\in T\subseteq \reals$. Then, for $U\subseteq T$ and
  $G,Q:S\to\bools$,
\begin{equation}
\left\{\lambda s\in S.\forall t\in U.\ \left(\forall
\tau\in {\downarrow}t.\ G\, (\flow_s\, \tau)\right) \rightarrow Q\,
(\flow_s\, t)\right\}\, x' = f\, \&\, G\, \{Q\}. \label{eq:h-evl}\tag{h-evl}
\end{equation}
\end{lemma}

This finishes the derivation of rules for a Hoare logic $\dH$ for
hybrid programs---to our knowledge, the first Hoare logic of this
kind. As usual, there is one rule per programming construct, so that
their recursive application generates proof obligations that are
entirely about data-level relationships---the discrete and continuous
evolution of hybrid program stores.

The rule~(\ref{eq:h-evl}) supports the following procedure
for reasoning with an evolution command $x' = f\, \&\, G$ and set $U$
in $\dH$:
\begin{enumerate}
\item Check that $f$ satisfies the conditions for Picard-Lindel\"of's
  theorem ($f$ is Lipschitz continuous and $S\subseteq\reals^n$ is
  open).
\item Supply a (local) flow $\flow$ for $f$ with open interval of
  existence $T$ around $0$.
\item Check that $\flow_s$ solves the IVP $(f,s)$ for each $s\in S$;
  ($\flow_s'\, t = f\, (\flow_s\, t)$, $\flow_s\, 0 = s$, and
  $U\subseteq T$).
\item If successful, apply rule~(\ref{eq:h-evl}).
\end{enumerate}

\begin{example}[Thermostat verification via solutions]\label{ex:therm-sol}
  A thermostat regulates the temperature $T$ of a room between bounds
  $T_l\le T\le T_h$. Variable $T_0$ stores an initial temperature;
  $\vartheta$ indicates whether the heater is switched on or off.
  Within time intervals of at most $\tau$ minutes, the thermostat
  resets time to $0$, measures the temperature, and turns the heater
  on or off dependent on the value obtained.  With $0<T_l$, $T_h<T_u$,
  $0 < a$,
  $U={\isacharbraceleft}{\isadigit{0}}{\isachardot}{\isachardot}{\isasymtau}{\isacharbraceright}=[0,\tau]$
  we define $f$, for $c\in\{0,T_u\}$, as

\begin{isabellebody}
\isanewline
\isacommand{abbreviation}\
{\isachardoublequoteopen}f\ a\ c\ {\isasymequiv}\ {\isacharbrackleft}T\ {\isasymmapsto}\isactrlsub s\ {\isacharminus}\ {\isacharparenleft}a\ {\isacharasterisk}\ {\isacharparenleft}T\ {\isacharminus}\ c{\isacharparenright}{\isacharparenright}{\isacharcomma}\ $T_0$\ {\isasymmapsto}\isactrlsub s\ {\isadigit{0}}{\isacharcomma}\ {\isasymtheta}\ {\isasymmapsto}\isactrlsub s\ {\isadigit{0}}{\isacharcomma}\ t\ {\isasymmapsto}\isactrlsub s\ {\isadigit{1}}{\isacharbrackright}{\isachardoublequoteclose}\isanewline
\end{isabellebody} 
\noindent Working alternatively with $\textit{vec-lens}^n_k$ or
$\textit{eucl-lens}^n_k $, we write $;$ instead of $\cdot$ for
sequential composition and use a guard $G$ to restrict evolutions
between $T_l$ and $T_h$ by setting
\begin{equation*}
G\, T_l\, T_h\, a\, c = \left(t\leq -\frac{1}{a}\ln\left(\frac{c-\Delta_c}{c-T_0}\right)\right),
\end{equation*}
where $\Delta_c = T_l$ if $c=0$, and $\Delta_c = T_h$ if $c=T_u$. The
hybrid program $\isa{therm}\ T_l\ T_h\ a\ T_u$ below models the
behaviour of the thermostat. To simplify notation, we separate into
a loop invariant ($I$), discrete control ($ctrl$), and continuous
dynamics ($dyn$).

\begin{isabellebody}

\isanewline
\isacommand{abbreviation}\
{\isachardoublequoteopen}I\ $T_l$\ $T_h$\ {\isasymequiv}\ \isactrlU{\isacharparenleft}$T_l$\ {\isasymle}\ T\ {\isasymand}\ T\ {\isasymle}\ $T_h$\ {\isasymand}\ {\isacharparenleft}{\isasymtheta}\ {\isacharequal}\ {\isadigit{0}}\ {\isasymor}\ {\isasymtheta}\ {\isacharequal}\ {\isadigit{1}}{\isacharparenright}{\isacharparenright}{\isachardoublequoteclose}\isanewline
\isanewline
\isacommand{abbreviation}\
{\isachardoublequoteopen}ctrl\ $T_l$\ $T_h$\ {\isasymequiv}\isanewline 
\ \ {\isacharparenleft}t\ {\isacharcolon}{\isacharcolon}{\isacharequal}\ {\isadigit{0}}{\isacharparenright}{\isacharsemicolon}\ {\isacharparenleft}$T_0$\ {\isacharcolon}{\isacharcolon}{\isacharequal}\ T{\isacharparenright}{\isacharsemicolon}\isanewline
\ \ {\isacharparenleft}IF\ {\isacharparenleft}{\isasymtheta}\ {\isacharequal}\ {\isadigit{0}}\ {\isasymand}\ $T_0$\ {\isasymle}\ $T_l$\ {\isacharplus}\ {\isadigit{1}}{\isacharparenright}\ THEN\ {\isacharparenleft}{\isasymtheta}\ {\isacharcolon}{\isacharcolon}{\isacharequal}\ {\isadigit{1}}{\isacharparenright}\ ELSE\ \isanewline
\ \ \ IF\ {\isacharparenleft}{\isasymtheta}\ {\isacharequal}\ {\isadigit{1}}\ {\isasymand}\ $T_0$\ {\isasymge}\ $T_h$\ {\isacharminus}\ {\isadigit{1}}{\isacharparenright}\ THEN\ {\isacharparenleft}{\isasymtheta}\ {\isacharcolon}{\isacharcolon}{\isacharequal}\ {\isadigit{0}}{\isacharparenright}\ ELSE\ skip{\isacharparenright}{\isachardoublequoteclose}\isanewline
\isanewline
\isacommand{abbreviation}\
{\isachardoublequoteopen}dyn\ $T_l$\ $T_h$\ a\ $T_u$\ {\isasymtau}\ {\isasymequiv}\ \isanewline
\ \ IF\ {\isacharparenleft}{\isasymtheta}\ {\isacharequal}\ {\isadigit{0}}{\isacharparenright}\ THEN\ x{\isasymacute}{\isacharequal}\ f\ a\ {\isadigit{0}}\ {\isacharampersand}\ G\ $T_l$\ $T_h$\ a\ {\isadigit{0}}\ on\ {\isacharbraceleft}{\isadigit{0}}{\isachardot}{\isachardot}{\isasymtau}{\isacharbraceright}\ UNIV\ {\isacharat}\ {\isadigit{0}}\ \isanewline
\ \ \ ELSE\ x{\isasymacute}{\isacharequal}\ f\ a\ $T_u$\ {\isacharampersand}\ G\ $T_l$\ $T_h$\ a\ $T_u$\ on\ {\isacharbraceleft}{\isadigit{0}}{\isachardot}{\isachardot}{\isasymtau}{\isacharbraceright}\ UNIV\ {\isacharat}\ {\isadigit{0}}{\isachardoublequoteclose}\isanewline
\isanewline
\isacommand{abbreviation}\ {\isachardoublequoteopen}therm\ $T_l$\ $T_h$\ a\ $T_u$\ {\isasymtau}\ {\isasymequiv}\isanewline 
\ \ LOOP\ {\isacharparenleft}ctrl\ $T_l$\ $T_h${\isacharsemicolon}\ dyn\ $T_l$\ $T_h$\ a\ $T_u$\ {\isasymtau}{\isacharparenright}\ INV\ {\isacharparenleft}I\ $T_l$\ $T_h${\isacharparenright}{\isachardoublequoteclose}\isanewline
\end{isabellebody}

\noindent The correctness specification and verification of the
thermostat with $\dH$  is then
\begin{isabellebody}
\isanewline
\isacommand{lemma}\isamarkupfalse%
\ thermostat{\isacharunderscore}flow{\isacharcolon}\ \isanewline
\ \ \isakeyword{assumes}\ {\isachardoublequoteopen}{\isadigit{0}}\ {\isacharless}\ a{\isachardoublequoteclose}\ \isakeyword{and}\ {\isachardoublequoteopen}{\isadigit{0}}\ {\isasymle}\ {\isasymtau}{\isachardoublequoteclose}\ \isakeyword{and}\ {\isachardoublequoteopen}{\isadigit{0}}\ {\isacharless}\ $T_l${\isachardoublequoteclose}\ \isakeyword{and}\ {\isachardoublequoteopen}$T_h$\ {\isacharless}\ $T_u${\isachardoublequoteclose}\isanewline
\ \ \isakeyword{shows}\ {\isachardoublequoteopen}\ \isactrlbold {\isacharbraceleft}I\ $T_l$\ $T_h$\isactrlbold {\isacharbraceright}\ therm\ $T_l$\ $T_h$\ a\ $T_u$\ {\isasymtau}\ \isactrlbold {\isacharbraceleft}I\ $T_l$\ $T_h$\isactrlbold {\isacharbraceright}{\isachardoublequoteclose}\isanewline
\ \ \isacommand{apply}\isamarkupfalse%
{\isacharparenleft}hyb{\isacharunderscore}hoare\ {\isachardoublequoteopen}{\isactrlU}{\isacharparenleft}I\ $T_l$\ $T_h$\ {\isasymand}\ t{\isacharequal}{\isadigit{0}}\ {\isasymand}\ T\isactrlsub {\isadigit{0}}\ {\isacharequal}\ T{\isacharparenright}{\isachardoublequoteclose}{\isacharparenright}\isanewline
\ \ \isacommand{prefer}\isamarkupfalse%
\ {\isadigit{4}}\ \isacommand{prefer}\isamarkupfalse%
\ {\isadigit{8}}\ \isacommand{using}\isamarkupfalse%
\ local{\isacharunderscore}flow{\isacharunderscore}therm\ assms\ \isacommand{apply}\isamarkupfalse\ force{\isacharplus}\isanewline
\ \ \isacommand{using}\isamarkupfalse%
\ assms\ therm{\isacharunderscore}dyn{\isacharunderscore}up\ therm{\isacharunderscore}dyn{\isacharunderscore}down\ \isacommand{by}\isamarkupfalse%
\ rel{\isacharunderscore}auto{\isacharprime}\isanewline
\end{isabellebody}

\noindent The first line uses tactic \isa{hyb-hoare} to blast away the
structure of \isa{therm} using $\dH$. To apply \isa{hyb-hoare}, the
program must be an iteration of the composition of two
programs---usually control and dynamics. The tactic requires lifting
the store to an Isabelle/UTP expression~\cite{FosterZW16}, which is
denoted by the $\isactrlU$ operator. Lemma \isa{local-flow-therm},
whose proof captures the procedure described above, supplies the flow
for $f\, a\, c$:
$\flow\, a\, c\, \tau = (-e^{-a\cdot\tau}(c-T)+c, \tau+t, T_0,
\vartheta)^\top$,
%\begin{isabellebody}
%\isanewline
%\isacommand{abbreviation}\isamarkupfalse%
%\ therm{\isacharunderscore}flow\ {\isacharcolon}{\isacharcolon}\ {\isachardoublequoteopen}real\ {\isasymRightarrow}\ real\ {\isasymRightarrow}\ real\ {\isasymRightarrow}\ {\isacharparenleft}real{\isacharcircum}{\isadigit{4}}{\isacharparenright}\ usubst{\isachardoublequoteclose}\ {\isacharparenleft}{\isachardoublequoteopen}{\isasymphi}{\isachardoublequoteclose}{\isacharparenright}\ \isanewline
%\ \ \isakeyword{where}\ {\isachardoublequoteopen}{\isasymphi}\ a\ $T_u$\ {\isasymtau}\ {\isasymequiv}\ {\isacharbrackleft}T\ {\isasymmapsto}\ {\isacharminus}\ exp{\isacharparenleft}{\isacharminus}a\ {\isacharasterisk}\ {\isasymtau}{\isacharparenright}\ {\isacharasterisk}\ {\isacharparenleft}L\ {\isacharminus}\ T{\isacharparenright}\ {\isacharplus}\ $T_u${\isacharcomma}\ t\ {\isasymmapsto}\ {\isasymtau}\ {\isacharplus}\ t{\isacharcomma}\ $T_0$\ {\isasymmapsto}\ $T_0${\isacharcomma}\ {\isasymtheta}\ {\isasymmapsto}\ {\isasymtheta}{\isacharbrackright}{\isachardoublequoteclose}\isanewline
%\isanewline
%\end{isabellebody}
for all $\tau\in\reals$. The remaining proof obligations are
inequalities of transcendental functions. They are discharged
automatically using auxiliary lemmas. \qed
\end{example}

%%%%%%%%%%%%%%%%%%%%%%%%%%%%%%%%%%%%%%%%%%%%%%%%%%

\section{Differential Hoare Logic for Invariants}\label{sec:hoare-inv}

Alternatively, $\dH$ supports reasoning with invariants for evolution
commands instead of supplying flows to~(\ref{eq:h-evl}).  The approach
has been developed in~\cite{MuniveS19}. Our invariants generalise the
\emph{differential invariants} of $\dL$~\cite{Platzer18} and the
\emph{invariant sets} of dynamical systems and (semi)group
theory~\cite{Teschl12} .

A predicate $I:S\to\bools$ is an \emph{invariant} of the continuous
vector field $f:S\to S$ and guard $G:S\to\bools$ \emph{along}
$T\subseteq \reals$ if
\begin{equation*}
\bigcup \Pow\, \gamma^f_G\, I\subseteq  I.
\end{equation*}
The operation $\bigcup\circ\Pow$ is the Kleisli extension $(-)^\dagger$
in the powerset monad. Hence we could simply write
$(\gamma^f_G)^\dagger\, I \subseteq I$. The definition of invariance
unfolds to
\begin{equation*}
  \forall s.\ I\, s \to (\forall X\in\Sols f\, T\, s.\forall t\in T.\ (\forall \tau\in {\downarrow}t.\ G\, (X\, \tau)) \to I\, (X\, t)).
\end{equation*}
For $G=\top$ we call $I$ an \emph{invariant} of $f$ along $T$.
Intuitively, invariants can be seen as sets of orbits. They are
coherent with the invariants from
Section~\ref{sec:hl-invariants}.
\begin{proposition}\label{P:inv-prop}
  Let $f:S\to S$ be continuous, $G:S\to\bools$ and
  $T\subseteq \reals$. Then $I$ is an invariant for $f$ and $G$ \emph{along} $T$ if and only if
$ \{I\}\, x' = f\, \&\, G\, \{I\}$.
\end{proposition}
Hence we can use a variant of (\ref{eq:h-inv}) for verification condition generation:
\begin{align}
  P\le I \land \{I\}\, x' = f\, \&\, G\, \{I\}\land (I\cdot G)\le Q\ \rightarrow\
  &\{P\}\, x' = f\, \&\, G\, \{Q\}.\label{eq:h-invg}\tag{h-invg}
\end{align}
It remains to check invariance in the antecedent of this rule. The
following lemma leads to a procedure.

\begin{lemma}[\cite{MuniveS19}]\label{P:invrules}
  Let $f:S\to S$ be a continuous vector field, $\mu,\nu:S\to\reals$
  differentiable and $T\subseteq \reals$. 
\begin{enumerate}
\item If $(\mu\circ X)' =(\nu\circ X)'$ for all
  $X\in \Sols f\, T\, s$, then $\{\mu=\nu\}\, x' = f\, \&\, G\, \{\mu=\nu\}$, 
\item if $(\mu\circ X)'\, t\leq(\nu\circ X)'\, t$ when $t> 0$, and $(\mu\circ X)'\, t\geq(\nu\circ X)'\, t$ when $t< 0$, for all $X\in \Sols f\, T\, s$,
  then $\{\mu <\nu\}\, x' = f\, \&\, G\, \{\mu <\nu\}$
\item $\mu\neq \nu$ if and only if $\mu < \nu$ or $\nu < \mu$,
\item $\mu \not\le \nu$ if and only if $\nu < \mu$.
\end{enumerate}
\end{lemma}

Condition $(1)$ follows from the well known fact that two continuously
differentiable functions are equal if they intersect at some point and
their derivatives are equal. Rules (\ref{eq:h-invg}),
(\ref{eq:h-inv-mult}), (\ref{eq:h-inv-plus}),
Proposition~\ref{P:inv-prop} and Lemma~\ref{P:invrules} yield the
following procedure for verifying $\{P\}\, x' = f\, \&\, G\, \{Q\}$:
\begin{enumerate}
\item Check whether a candidate predicate $I$ is an invariant for $f$
  along $T$:
	\begin{enumerate}
	\item transform $I$ into negation normal form;
	\item reduce complex $I$ (with (\ref{eq:h-inv-mult}), (\ref{eq:h-inv-plus}) and Lemma~\ref{P:invrules} (3,4);
	\item if $I$ is atomic, apply Lemma~\ref{P:invrules} (1) and (2);
	\end{enumerate}
(if successful,  $\{I\}\, x' = f\, \&\, G\, \{I\}$ holds by Proposition~\ref{P:inv-prop}),
\item if successful, prove $P\le I$ and $(I\cdot G)\le Q$ to apply rule (\ref{eq:h-invg}).
\end{enumerate}

\begin{example}[Water tank verification via invariants]\label{ex:tank-inv}
  A controller turns a water pump on and off to keep the water level
  $h$ in a tank within safe bounds $h_l\leq h\leq h_h$.  Variable
  $h_0$ stores an initial water level;  $\pi$ indicates whether the pump is
  on or off. The rate of change of the water-level is
  linear with slope $k\in\{-c_o,c_i-c_o\}$ (assuming $c_i>c_o$). The
  vector field \isa{f} for this behaviour and its invariant \isa{dI} are
\begin{isabellebody}
\isanewline
\isacommand{abbreviation}\
{\isachardoublequoteopen}f\ k\ {\isasymequiv}\ {\isacharbrackleft}{\isasympi}\ {\isasymmapsto}\isactrlsub s\ {\isadigit{0}}{\isacharcomma}\ h\ {\isasymmapsto}\isactrlsub s\ k{\isacharcomma}\ $h_0$\ {\isasymmapsto}\isactrlsub s\ {\isadigit{0}}{\isacharcomma}\ t\ {\isasymmapsto}\isactrlsub s\ {\isadigit{1}}{\isacharbrackright}{\isachardoublequoteclose}\isanewline
\isanewline
\isacommand{abbreviation}\ 
{\isachardoublequoteopen}dI\ $h_l$\ $h_h$\ k\ {\isasymequiv}\isanewline
\ \ {\isactrlU}{\isacharparenleft}h\ {\isacharequal}\ k\ {\isasymcdot}\ t\
{\isacharplus}\ $h_0$\ {\isasymand}\ {\isadigit{0}}\ {\isasymle}\ t\
{\isasymand}\ $h_l$\ {\isasymle}\ $h_0$\ {\isasymand}\ $h_0$\
{\isasymle}\ $h_h$\ {\isasymand}\ {\isacharparenleft}{\isasympi}\
{\isacharequal}\ {\isadigit{0}}\ {\isasymor}\ {\isasympi}\
{\isacharequal}\
{\isadigit{1}}{\isacharparenright}{\isacharparenright}{\isachardoublequoteclose}
\isanewline
\end{isabellebody}
\noindent This vector field differs from that of the thermostat
(Example~\ref{ex:therm-sol}). A hybrid program for the controller is
given once again by guard $G\ h_x\ k$ with
$h_x\in\{h_l,h_h\}$ that restricts evolutions beyond $h_x$,
loop invariant $I$, control and dynamic part before the program
itself:
\begin{isabellebody}
\isanewline
\isacommand{abbreviation}\ 
{\isachardoublequoteopen}G\ h\isactrlsub x\ k\ {\isasymequiv}\ {\isactrlU}{\isacharparenleft}t\ {\isasymle}\ {\isacharparenleft}h\isactrlsub x\ {\isacharminus}\ $h_0${\isacharparenright}{\isacharslash}k{\isacharparenright}{\isachardoublequoteclose}\isanewline
\isanewline
\isacommand{abbreviation}\
{\isachardoublequoteopen}I\ $h_l$\ $h_h$\ {\isasymequiv}\ {\isactrlU}{\isacharparenleft}$h_l$\ {\isasymle}\ h\ {\isasymand}\ h\ {\isasymle}\ $h_h$\ {\isasymand}\ {\isacharparenleft}{\isasympi}\ {\isacharequal}\ {\isadigit{0}}\ {\isasymor}\ {\isasympi}\ {\isacharequal}\ {\isadigit{1}}{\isacharparenright}{\isacharparenright}{\isachardoublequoteclose}\isanewline
\isanewline
\isacommand{abbreviation}\
{\isachardoublequoteopen}ctrl\ $h_l$\ $h_h$\ {\isasymequiv}\isanewline
\ \ {\isacharparenleft}t\ {\isacharcolon}{\isacharcolon}{\isacharequal}{\isadigit{0}}{\isacharparenright}{\isacharsemicolon}{\isacharparenleft}$h_0$\ {\isacharcolon}{\isacharcolon}{\isacharequal}\ h{\isacharparenright}{\isacharsemicolon}\isanewline
\ \ {\isacharparenleft}IF\ {\isacharparenleft}{\isasympi}\ {\isacharequal}\ {\isadigit{0}}\ {\isasymand}\ $h_0$\ {\isasymle}\ $h_l$\ {\isacharplus}\ {\isadigit{1}}{\isacharparenright}\ THEN\ {\isacharparenleft}{\isasympi}\ {\isacharcolon}{\isacharcolon}{\isacharequal}\ {\isadigit{1}}{\isacharparenright}\ ELSE\isanewline
\ \ {\isacharparenleft}IF\ {\isacharparenleft}{\isasympi}\ {\isacharequal}\ {\isadigit{1}}\ {\isasymand}\ $h_0$\ {\isasymge}\ $h_h$\ {\isacharminus}\ {\isadigit{1}}{\isacharparenright}\ THEN\ {\isacharparenleft}{\isasympi}\ {\isacharcolon}{\isacharcolon}{\isacharequal}\ {\isadigit{0}}{\isacharparenright}\ ELSE\ skip{\isacharparenright}{\isacharparenright}{\isachardoublequoteclose}\isanewline
\isanewline
\isacommand{abbreviation}\
{\isachardoublequoteopen}dyn\ c\isactrlsub i\ c\isactrlsub o\ $h_l$\ $h_h$\ {\isasymtau}\ {\isasymequiv}\ IF\ {\isacharparenleft}{\isasympi}\ {\isacharequal}\ {\isadigit{0}}{\isacharparenright}\ THEN\ \isanewline
\ \ \ \ x{\isasymacute}{\isacharequal}\ f\ {\isacharparenleft}c\isactrlsub i{\isacharminus}c\isactrlsub o{\isacharparenright}\ {\isacharampersand}\ G\ $h_h$\ {\isacharparenleft}c\isactrlsub i{\isacharminus}c\isactrlsub o{\isacharparenright}\ on\ {\isacharbraceleft}{\isadigit{0}}{\isachardot}{\isachardot}{\isasymtau}{\isacharbraceright}\ UNIV\ {\isacharat}\ {\isadigit{0}}\ DINV\ {\isacharparenleft}dI\ $h_l$\ $h_h$\ {\isacharparenleft}c\isactrlsub i{\isacharminus}c\isactrlsub o{\isacharparenright}{\isacharparenright}\isanewline
\ \ ELSE\ x{\isasymacute}{\isacharequal}\ f\ {\isacharparenleft}{\isacharminus}c\isactrlsub o{\isacharparenright}\ {\isacharampersand}\ G\ $h_l$\ {\isacharparenleft}{\isacharminus}c\isactrlsub o{\isacharparenright}\ on\ {\isacharbraceleft}{\isadigit{0}}{\isachardot}{\isachardot}{\isasymtau}{\isacharbraceright}\ UNIV\ {\isacharat}\ {\isadigit{0}}\ DINV\ {\isacharparenleft}dI\ $h_l$\ $h_h$\ {\isacharparenleft}{\isacharminus}c\isactrlsub o{\isacharparenright}{\isacharparenright}{\isachardoublequoteclose}\isanewline
\isanewline
\isacommand{abbreviation}\isamarkupfalse%
\ {\isachardoublequoteopen}tank{\isacharunderscore}dinv\ c\isactrlsub i\ c\isactrlsub o\ $h_l$\ $h_h$\ {\isasymtau}\ {\isasymequiv}\isanewline 
\ \ LOOP\ {\isacharparenleft}ctrl\ $h_l$\ $h_h${\isacharsemicolon}\ dyn\ c\isactrlsub i\ c\isactrlsub o\ $h_l$\ $h_h$\ {\isasymtau}{\isacharparenright}\ INV\ {\isacharparenleft}I\ $h_l$\ $h_h${\isacharparenright}{\isachardoublequoteclose}\isanewline
\end{isabellebody}

\noindent The correctness specification and verification of the water
tank with $\dH$ is then

\begin{isabellebody}
\isanewline
\isacommand{lemma}\isamarkupfalse%
\ tank{\isacharunderscore}diff{\isacharunderscore}inv{\isacharcolon}\ {\isachardoublequoteopen}{\isadigit{0}}\ {\isasymle}\ {\isasymtau}\ {\isasymLongrightarrow}\ diff{\isacharunderscore}invariant\ {\isacharparenleft}dI\ $h_l$\ $h_h$\ k{\isacharparenright}\ {\isacharparenleft}f\ k{\isacharparenright}\ {\isacharbraceleft}{\isadigit{0}}{\isachardot}{\isachardot}{\isasymtau}{\isacharbraceright}\ UNIV\ {\isadigit{0}}\ Guard{\isachardoublequoteclose}\isanewline
\ \ $\langle \isa{proof}\rangle$\isanewline
\isanewline
\isacommand{lemma}\isamarkupfalse%
\ tank{\isacharunderscore}inv{\isacharcolon}\isanewline
\ \ \isakeyword{assumes}\ {\isachardoublequoteopen}{\isadigit{0}}\ {\isasymle}\ {\isasymtau}{\isachardoublequoteclose}\ \isakeyword{and}\ {\isachardoublequoteopen}{\isadigit{0}}\ {\isacharless}\ c\isactrlsub o{\isachardoublequoteclose}\ \isakeyword{and}\ {\isachardoublequoteopen}c\isactrlsub o\ {\isacharless}\ c\isactrlsub i{\isachardoublequoteclose}\isanewline
\ \ \isakeyword{shows}\ {\isachardoublequoteopen}\ \isactrlbold {\isacharbraceleft}I\ $h_l$\ $h_h$\isactrlbold {\isacharbraceright}\ tank{\isacharunderscore}dinv\ c\isactrlsub i\ c\isactrlsub o\ $h_l$\ $h_h$\ {\isasymtau}\ \isactrlbold {\isacharbraceleft}I\ $h_l$\ $h_h$\isactrlbold {\isacharbraceright}{\isachardoublequoteclose}\isanewline
\ \ \isacommand{apply}\isamarkupfalse%
{\isacharparenleft}hyb{\isacharunderscore}hoare\ {\isachardoublequoteopen}{\isactrlU}{\isacharparenleft}I\ $h_l$\ $h_h$\ {\isasymand}\ t\ {\isacharequal}\ {\isadigit{0}}\ {\isasymand}\ h\isactrlsub {\isadigit{0}}\ {\isacharequal}\ h{\isacharparenright}{\isachardoublequoteclose}{\isacharparenright}\isanewline
\ \ \isacommand{prefer}\isamarkupfalse%
\ {\isadigit{4}}\ \isacommand{prefer}\isamarkupfalse%
\ {\isadigit{7}}\ \isacommand{using}\isamarkupfalse%
\ tank{\isacharunderscore}diff{\isacharunderscore}inv\ assms\ \isacommand{apply}\isamarkupfalse%
\ force{\isacharplus}\isanewline
\ \ \isacommand{using}\isamarkupfalse%
\ assms\ tank{\isacharunderscore}inv{\isacharunderscore}arith{\isadigit{1}}\ tank{\isacharunderscore}inv{\isacharunderscore}arith{\isadigit{2}}\ \isacommand{by}\isamarkupfalse%
\ rel{\isacharunderscore}auto{\isacharprime}\isanewline
\end{isabellebody}

\noindent As in Example~\ref{ex:therm-sol}, tactic \isa{hyb-hoare}
blasts away the control structure of this program. The second proof
line uses Lemma \isa{tank-diff-inv} to check that \isa{dI} is an
invariant, using the procedure outlined (see our repository for
technical details). Auxiliary lemmas then discharge the remaining
arithmetic proof obligations.  \qed
\end{example}

%%%%%%%%%%%%%%%%%%%%%%%%%%%%%%%%%%%%%%%%%%%%%%%%%%

\section{Differential Refinement Calculi}\label{sec:refine}

A \emph{refinement Kleene algebra with tests}
($\rKAT$)~\cite{ArmstrongGS16}  is a $\KAT$
$(K,B)$ expanded by an operation $[-,-]:B\times B\to K$ that
satisfies, for all $\alpha \in K$ and $p,q\in B$, 
\begin{equation*}
  \{p\}\, \alpha\, \{q\} \leftrightarrow \alpha\le [p,q].
\end{equation*}
The element $[p,q]$ of $K$ corresponds to Morgan's \emph{specification
  statement}~\cite{Morgan94}. It satisfies $\{p\}\, [p,q]\, \{q\}$
and $\{p\}\, \alpha\, \{q\} \rightarrow \alpha\le [p,q]$, 
which makes $[p,q]$ the greatest element of $K$ that satisfies the Hoare
triple with precondition $p$ and postcondition $q$.  Indeed, in
$\sta\, S$ and for $S\subseteq \reals^n$,
$ [P,Q] = \bigcup \left\{f:S\to \Pow\, S \mid \{P\}\, f\,
  \{Q\}\right\}$. 

Variants of Morgan's laws~\cite{Morgan94} of a \emph{propositional
  refinement calculus}---once more ignoring assignments---are then derivable in
$\rKAT$~\cite{ArmstrongGS16}.
\begin{align}
  1 &\le [p,p],\label{eq:r-skip}\tag{r-skip}\\
[p',q'] &\le [p,q],\qquad \text{ if } p\le p'\text{ and } q'\le q,\label{eq:r-cons}\tag{r-cons}\\
[p,r]\cdot [r,q] &\le [p,q],\label{eq:r-seq}\tag{r-seq}\\
\IF{t}{[tp,q]}{[\neg tp,q]} &\le [p,q],\label{eq:r-cond}\tag{r-cond}\\
 \WHILE{t}{[tp,p]} &\le [p,\neg tp]. \label{eq:r-while}\tag{r-while}
\end{align}
We have also derived $\alpha \le [0,1]$ and $[1,0] \le \alpha$, but do
not use them in proofs.

For invariants and loops, we obtain the additional refinement laws
\begin{align}
  [i,i] &\le [p,q],\qquad \text{ if } p\le i \le q,\label{eq:r-inv}\tag{r-inv}\\
\mathbf{loop}\, [i,i] &\le [i,i]. \label{eq:r-loop}\tag{r-loop}
\end{align}

In $\sta\, S$, moreover, the following assignments laws are
derivable~\cite{ArmstrongGS16}.
\begin{align}
 (x := e)  &\le  \left[Q[e/x],Q\right],\label{eq:r-assgn}\tag{r-assgn}\\
(x:= e) \cdot \left[Q,Q\right] &\le [Q[e/x],Q],\label{eq:r-assgn}\tag{r-assgnl}\\
\left[Q,Q[e/x]\right]\cdot (x:=e) &\le [Q,Q]. \label{eq:r-assgn}\tag{r-assgnf}
\end{align}
The second and third law are known as \emph{leading} and \emph{following}
law. They introduce an assignment before and after a block of code. 

Finally, we obtain the following refinement laws for evolution
commands.
\begin{lemma}\label{P:r-evl-lemma}
  Let $f:S\to S$ be a Lipschitz continuous vector field on
  $S\subseteq \reals^n$ and $\flow:T\to S\to S$ its local flow with
  $0\in T\subseteq \reals$. Then, for $U\subseteq T$ and
  $G,Q:S\to\bools$,
\begin{gather*}
(x' = f\, \&\, G)\, \le\, [\lambda s.\forall t\in U.\ (\forall
\tau\in {\downarrow}t.\ G\, (\flow_s\, \tau))\to Q\, (\flow_s\, t),Q],\label{eq:r-evl}\tag{r-evl}\\
(x' = f\, \&\, G) \cdot \left[Q,Q\right]\, \le\, [\lambda s. \forall t\in U.\ (\forall
\tau\in {\downarrow}t.\ G\, (\flow_s\, \tau))\to Q\, (\flow_s\, t),Q],\label{eq:r-evll}\tag{r-evll}\\
\left[Q,\lambda s. \forall t\in U.\ (\forall
\tau\in {\downarrow}t.\ G\, (\flow_s\, \tau))\to Q\, (\flow_s\,
t)\right]\cdot (x' = f\, \&\, G) \, \le \, [Q,Q].\label{eq:r-evlr}\tag{r-evlr}
\end{gather*}
\end{lemma}

The laws in this section form the differential refinement calculus
$\dR$.  They suffice for constructing hybrid programs from initial
specification statements by step-wise refinement incrementally and
compositionally. To our knowledge, $\dR$ is the first refinement
calculus for hybrid programs of this kind.  A more powerful variant
based on predicate transformers \`a la Back and von
Wright~\cite{BackW98} has been developed in~\cite{MuniveS19}, but
applications remain to be explored.  A previous approach to refinement
for hybrid programs in $\dL$~\cite{LoosP16} is quite different to the
two standard calculi mentioned and much more intricate than the
approach presented.

\begin{example}[Thermostat refinement via solutions]\label{ex:therm-rsol}
  We now construct program \isa{therm} from Example~\ref{ex:therm-sol}
  by step-wise refinement using the rules of $\dR$. 
%\begin{equation*}
%\isa{\isactrlbold {\isacharbrackleft}I\ $T_l$\ $T_h${\isacharcomma}\ I\ $T_l$\ $T_h$\isactrlbold {\isacharbrackright}\ {\isasymge}\ therm\ $T_l$\ $T_h$\ a\ $T_u$\ {\isasymtau}}.
%\end{equation*} 
%First we refine hybrid programs for the discrete control and the
%continuous dynamics separately from specification statements. Then we
%combine them into a refinement statement of the loop, and finally we
%add refinements for the variable initialisations to construct the
%entire program from its specification.  

\begin{isabellebody}
\isanewline
\isacommand{lemma}\isamarkupfalse%
\ R{\isacharunderscore}therm{\isacharunderscore}down{\isacharcolon}\ \isanewline
\ \ \isakeyword{assumes}\ {\isachardoublequoteopen}a\ {\isachargreater}\ {\isadigit{0}}{\isachardoublequoteclose}\ \isakeyword{and}\ {\isachardoublequoteopen}{\isadigit{0}}\ {\isasymle}\ {\isasymtau}{\isachardoublequoteclose}\ \isakeyword{and}\ {\isachardoublequoteopen}{\isadigit{0}}\ {\isacharless}\ T\isactrlsub l{\isachardoublequoteclose}\ \isakeyword{and}\ {\isachardoublequoteopen}T\isactrlsub h\ {\isacharless}\ T\isactrlsub u{\isachardoublequoteclose}\isanewline
\ \ \isakeyword{shows}\ {\isachardoublequoteopen}\isactrlbold {\isacharbrackleft}{\isasymtheta}\ {\isacharequal}\ {\isadigit{0}}\ {\isasymand}\ I\ T\isactrlsub l\ T\isactrlsub h\ {\isasymand}\ t\ {\isacharequal}\ {\isadigit{0}}\ {\isasymand}\ T\isactrlsub {\isadigit{0}}\ {\isacharequal}\ T{\isacharcomma}\ I\ T\isactrlsub l\ T\isactrlsub h\isactrlbold {\isacharbrackright}\ {\isasymge}\isanewline 
\ \ {\isacharparenleft}x{\isasymacute}{\isacharequal}\ f\ a\ {\isadigit{0}}\ {\isacharampersand}\ G\ T\isactrlsub l\ T\isactrlsub h\ a\ {\isadigit{0}}\ on\ {\isacharbraceleft}{\isadigit{0}}{\isachardot}{\isachardot}{\isasymtau}{\isacharbraceright}\ UNIV\ {\isacharat}\ {\isadigit{0}}{\isacharparenright}{\isachardoublequoteclose}\isanewline
\ \ \isacommand{apply}\isamarkupfalse%
{\isacharparenleft}rule\ local{\isacharunderscore}flow{\isachardot}R{\isacharunderscore}g{\isacharunderscore}ode{\isacharunderscore}ivl{\isacharbrackleft}OF\ local{\isacharunderscore}flow{\isacharunderscore}therm{\isacharbrackright}{\isacharparenright}\isanewline
\ \ \isacommand{using}\isamarkupfalse%
\ therm{\isacharunderscore}dyn{\isacharunderscore}down{\isacharbrackleft}OF\ assms{\isacharparenleft}{\isadigit{1}}{\isacharcomma}{\isadigit{3}}{\isacharparenright}{\isacharcomma}\ of\ {\isacharunderscore}\ T\isactrlsub h{\isacharbrackright}\ assms\ \isacommand{by}\isamarkupfalse%
\ rel{\isacharunderscore}auto{\isacharprime}%
\isanewline
\isanewline
\isacommand{lemma}\isamarkupfalse%
\ R{\isacharunderscore}therm{\isacharunderscore}up{\isacharcolon}\ \isanewline
\ \ \isakeyword{assumes}\ {\isachardoublequoteopen}a\ {\isachargreater}\ {\isadigit{0}}{\isachardoublequoteclose}\ \isakeyword{and}\ {\isachardoublequoteopen}{\isadigit{0}}\ {\isasymle}\ {\isasymtau}{\isachardoublequoteclose}\ \isakeyword{and}\ {\isachardoublequoteopen}{\isadigit{0}}\ {\isacharless}\ T\isactrlsub l{\isachardoublequoteclose}\ \isakeyword{and}\ {\isachardoublequoteopen}T\isactrlsub h\ {\isacharless}\ T\isactrlsub u{\isachardoublequoteclose}\isanewline
\ \ \isakeyword{shows}\ {\isachardoublequoteopen}\isactrlbold {\isacharbrackleft}{\isasymnot}\ {\isasymtheta}\ {\isacharequal}\ {\isadigit{0}}\ {\isasymand}\ I\ T\isactrlsub l\ T\isactrlsub h\ {\isasymand}\ t\ {\isacharequal}\ {\isadigit{0}}\ {\isasymand}\ T\isactrlsub {\isadigit{0}}\ {\isacharequal}\ T{\isacharcomma}\ I\ T\isactrlsub l\ T\isactrlsub h\isactrlbold {\isacharbrackright}\ {\isasymge}\isanewline
\ \ {\isacharparenleft}x{\isasymacute}{\isacharequal}\ f\ a\ T\isactrlsub u\ {\isacharampersand}\ G\ T\isactrlsub l\ T\isactrlsub h\ a\ T\isactrlsub u\ on\ {\isacharbraceleft}{\isadigit{0}}{\isachardot}{\isachardot}{\isasymtau}{\isacharbraceright}\ UNIV\ {\isacharat}\ {\isadigit{0}}{\isacharparenright}{\isachardoublequoteclose}\isanewline
\ \ \isacommand{apply}\isamarkupfalse%
{\isacharparenleft}rule\ local{\isacharunderscore}flow{\isachardot}R{\isacharunderscore}g{\isacharunderscore}ode{\isacharunderscore}ivl{\isacharbrackleft}OF\ local{\isacharunderscore}flow{\isacharunderscore}therm{\isacharbrackright}{\isacharparenright}\isanewline
\ \ \isacommand{using}\isamarkupfalse%
\ therm{\isacharunderscore}dyn{\isacharunderscore}up{\isacharbrackleft}OF\ assms{\isacharparenleft}{\isadigit{1}}{\isacharparenright}\ {\isacharunderscore}\ {\isacharunderscore}\ assms{\isacharparenleft}{\isadigit{4}}{\isacharparenright}{\isacharcomma}\ of\ T\isactrlsub l{\isacharbrackright}\ assms\ \isacommand{by}\isamarkupfalse%
\ rel{\isacharunderscore}auto{\isacharprime}%
\isanewline
\isanewline
\isacommand{lemma}\isamarkupfalse%
\ R{\isacharunderscore}therm{\isacharunderscore}time{\isacharcolon}\ {\isachardoublequoteopen}\isactrlbold {\isacharbrackleft}I\ T\isactrlsub l\ T\isactrlsub h{\isacharcomma}\ I\ T\isactrlsub l\ T\isactrlsub h\ {\isasymand}\ t\ {\isacharequal}\ {\isadigit{0}}\isactrlbold {\isacharbrackright}\ {\isasymge}\ {\isacharparenleft}t\ {\isacharcolon}{\isacharcolon}{\isacharequal}\ {\isadigit{0}}{\isacharparenright}{\isachardoublequoteclose}\isanewline
\ \ \isacommand{by}\isamarkupfalse%
\ {\isacharparenleft}rule\ R{\isacharunderscore}assign{\isacharunderscore}law{\isacharcomma}\ pred{\isacharunderscore}simp{\isacharparenright}%
\isanewline
\isanewline
\isacommand{lemma}\isamarkupfalse%
\ R{\isacharunderscore}therm{\isacharunderscore}temp{\isacharcolon}\ {\isachardoublequoteopen}\isactrlbold {\isacharbrackleft}I\ T\isactrlsub l\ T\isactrlsub h\ {\isasymand}\ t\ {\isacharequal}\ {\isadigit{0}}{\isacharcomma}\ I\ T\isactrlsub l\ T\isactrlsub h\ {\isasymand}\ t\ {\isacharequal}\ {\isadigit{0}}\ {\isasymand}\ T\isactrlsub {\isadigit{0}}\ {\isacharequal}\ T\isactrlbold {\isacharbrackright}\ {\isasymge}\ {\isacharparenleft}T\isactrlsub {\isadigit{0}}\ {\isacharcolon}{\isacharcolon}{\isacharequal}\ T{\isacharparenright}{\isachardoublequoteclose}\isanewline
\ \ \isacommand{by}\isamarkupfalse%
\ {\isacharparenleft}rule\ R{\isacharunderscore}assign{\isacharunderscore}law{\isacharcomma}\ pred{\isacharunderscore}simp{\isacharparenright}%
\isanewline
\isanewline
\isacommand{lemma}\isamarkupfalse%
\ R{\isacharunderscore}thermostat{\isacharunderscore}flow{\isacharcolon}\isanewline
\ \ \isakeyword{assumes}\ {\isachardoublequoteopen}a\ {\isachargreater}\ {\isadigit{0}}{\isachardoublequoteclose}\ \isakeyword{and}\ {\isachardoublequoteopen}{\isadigit{0}}\ {\isasymle}\ {\isasymtau}{\isachardoublequoteclose}\ \isakeyword{and}\ {\isachardoublequoteopen}{\isadigit{0}}\ {\isacharless}\ T\isactrlsub l{\isachardoublequoteclose}\ \isakeyword{and}\ {\isachardoublequoteopen}T\isactrlsub h\ {\isacharless}\ T\isactrlsub u{\isachardoublequoteclose}\isanewline
\ \ \isakeyword{shows}\ {\isachardoublequoteopen}\isactrlbold {\isacharbrackleft}I\ T\isactrlsub l\ T\isactrlsub h{\isacharcomma}\ I\ T\isactrlsub l\ T\isactrlsub h\isactrlbold {\isacharbrackright}\ {\isasymge}\ therm\ T\isactrlsub l\ T\isactrlsub h\ a\ T\isactrlsub u\ {\isasymtau}{\isachardoublequoteclose}\isanewline
\ \ \isacommand{by}\isamarkupfalse%
\ {\isacharparenleft}refinement{\isacharsemicolon}{\isacharparenleft}rule\ R{\isacharunderscore}therm{\isacharunderscore}time{\isacharparenright}{\isacharquery}{\isacharcomma}{\isacharparenleft}rule\ R{\isacharunderscore}therm{\isacharunderscore}temp{\isacharparenright}{\isacharquery}{\isacharcomma}{\isacharparenleft}rule R{\isacharunderscore}assign{\isacharunderscore}law{\isacharparenright}{\isacharquery}{\isacharcomma} \isanewline
\ \ \ \ \ \ {\isacharparenleft}rule\ R{\isacharunderscore}therm{\isacharunderscore}up{\isacharbrackleft}OF\ assms{\isacharbrackright}{\isacharparenright}{\isacharquery}{\isacharcomma}\ {\isacharparenleft}rule\ R{\isacharunderscore}therm{\isacharunderscore}down{\isacharbrackleft}OF\ assms{\isacharbrackright}{\isacharparenright}{\isacharquery}{\isacharparenright}\ rel{\isacharunderscore}auto{\isacharprime}
\isanewline
\end{isabellebody}
The \isa{refinement} tactic pushes the refinement specification
through the program structure until the only remaining proof
obligations are atomic refinements. We only refine the atomic programs
needed to complete proofs automatically; 
those for the first two assignment and the evolution commands. \qed
\end{example}

\begin{example}[Water tank refinement via invariants]\label{ex:tank-rinv}
  Alternatively we may use differential invariants with $\dR$ to
  refine \isa{tank{\isacharunderscore}dinv} from
  Example~\ref{ex:tank-inv}. This time we supply a single structured
  proof to show another style of refinement.
  We %omit some parts of the proof and
  abbreviate long expressions with schematic variables.

\begin{isabellebody}
\isanewline
\isacommand{lemma}\isamarkupfalse%
\ R{\isacharunderscore}tank{\isacharunderscore}inv{\isacharcolon}\isanewline
\ \ \isakeyword{assumes}\ {\isachardoublequoteopen}{\isadigit{0}}\ {\isasymle}\ {\isasymtau}{\isachardoublequoteclose}\ \isakeyword{and}\ {\isachardoublequoteopen}{\isadigit{0}}\ {\isacharless}\ c\isactrlsub o{\isachardoublequoteclose}\ \isakeyword{and}\ {\isachardoublequoteopen}c\isactrlsub o\ {\isacharless}\ c\isactrlsub i{\isachardoublequoteclose}\isanewline
\ \ \isakeyword{shows}\ {\isachardoublequoteopen}\isactrlbold {\isacharbrackleft}I\ h\isactrlsub l\ h\isactrlsub h{\isacharcomma}\ I\ h\isactrlsub l\ h\isactrlsub h\isactrlbold {\isacharbrackright}\ {\isasymge}\ tank{\isacharunderscore}dinv\ c\isactrlsub i\ c\isactrlsub o\ h\isactrlsub l\ h\isactrlsub h\ {\isasymtau}{\isachardoublequoteclose}\isanewline
\isacommand{proof}\isamarkupfalse%
{\isacharminus}\isanewline
\ \ \isacommand{have}\isamarkupfalse%
\ {\isachardoublequoteopen}\isactrlbold {\isacharbrackleft}I\ h\isactrlsub l\ h\isactrlsub h{\isacharcomma}\ I\ h\isactrlsub l\ h\isactrlsub h\isactrlbold {\isacharbrackright}\ {\isasymge}\isanewline
\ \ \ LOOP\ {\isacharparenleft}{\isacharparenleft}t\ {\isacharcolon}{\isacharcolon}{\isacharequal}\ {\isadigit{0}}{\isacharparenright}{\isacharsemicolon}\isactrlbold {\isacharbrackleft}I\ h\isactrlsub l\ h\isactrlsub h\ {\isasymand}\ t\ {\isacharequal}\ {\isadigit{0}}{\isacharcomma}\ I\ h\isactrlsub l\ h\isactrlsub h\isactrlbold {\isacharbrackright}{\isacharparenright}\ INV\ I\ h\isactrlsub l\ h\isactrlsub h{\isachardoublequoteclose}\ {\isacharparenleft}\isakeyword{is}\ {\isachardoublequoteopen}{\isacharunderscore}\ {\isasymge}\ {\isacharquery}R{\isachardoublequoteclose}{\isacharparenright}\isanewline
\ \ \ \ \isacommand{by}\isamarkupfalse%
\ {\isacharparenleft}refinement{\isacharcomma}\ rel{\isacharunderscore}auto{\isacharprime}{\isacharparenright}\isanewline
\ \ \isacommand{moreover}\isamarkupfalse%
\ \isacommand{have}\isamarkupfalse%
\ {\isachardoublequoteopen}{\isacharquery}R\ {\isasymge}\ LOOP\isanewline
\ \ \ {\isacharparenleft}{\isacharparenleft}t\ {\isacharcolon}{\isacharcolon}{\isacharequal}\ {\isadigit{0}}{\isacharparenright}{\isacharsemicolon}{\isacharparenleft}h\isactrlsub {\isadigit{0}}\ {\isacharcolon}{\isacharcolon}{\isacharequal}\ h{\isacharparenright}{\isacharsemicolon}\isactrlbold {\isacharbrackleft}I\ h\isactrlsub l\ h\isactrlsub h\ {\isasymand}\ t\ {\isacharequal}\ {\isadigit{0}}\ {\isasymand}\ h\isactrlsub {\isadigit{0}}\ {\isacharequal}\ h{\isacharcomma}\ I\ h\isactrlsub l\ h\isactrlsub h\isactrlbold {\isacharbrackright}{\isacharparenright}\ INV\ I\ h\isactrlsub l\ h\isactrlsub h{\isachardoublequoteclose}\ {\isacharparenleft}\isakeyword{is}\ {\isachardoublequoteopen}{\isacharunderscore}\ {\isasymge}\ {\isacharquery}R{\isachardoublequoteclose}{\isacharparenright}\isanewline
\ \ \ \ \isacommand{by}\isamarkupfalse%
\ {\isacharparenleft}refinement{\isacharcomma}\ rel{\isacharunderscore}auto{\isacharprime}{\isacharparenright}\isanewline
\ \ \isacommand{moreover}\isamarkupfalse%
\ \isacommand{have}\isamarkupfalse%
\ {\isachardoublequoteopen}{\isacharquery}R\ {\isasymge}\isanewline
\ \ \ LOOP\ {\isacharparenleft}ctrl\ h\isactrlsub l\ h\isactrlsub h{\isacharsemicolon}\isactrlbold {\isacharbrackleft}I\ h\isactrlsub l\ h\isactrlsub h\ {\isasymand}\ t\ {\isacharequal}\ {\isadigit{0}}\ {\isasymand}\ h\isactrlsub {\isadigit{0}}\ {\isacharequal}\ h{\isacharcomma}\ I\ h\isactrlsub l\ h\isactrlsub h\isactrlbold {\isacharbrackright}{\isacharparenright}\ INV\ I\ h\isactrlsub l\ h\isactrlsub h{\isachardoublequoteclose}\ {\isacharparenleft}\isakeyword{is}\ {\isachardoublequoteopen}{\isacharunderscore}\ {\isasymge}\ {\isacharquery}R{\isachardoublequoteclose}{\isacharparenright}\isanewline
\ \ \ \ \isacommand{by}\isamarkupfalse%
\ {\isacharparenleft}simp\ only{\isacharcolon}\ mult{\isachardot}assoc{\isacharcomma}\ refinement{\isacharsemicolon}\ {\isacharparenleft}force{\isacharparenright}{\isacharquery}{\isacharcomma}\ {\isacharparenleft}rule\ R{\isacharunderscore}assign{\isacharunderscore}law{\isacharparenright}{\isacharquery}{\isacharparenright}\ rel{\isacharunderscore}auto{\isacharprime}\isanewline
\ \ \isacommand{moreover}\isamarkupfalse%
\ \isacommand{have}\isamarkupfalse%
\ {\isachardoublequoteopen}{\isacharquery}R\ {\isasymge}\ LOOP\ {\isacharparenleft}ctrl\ h\isactrlsub l\ h\isactrlsub h{\isacharsemicolon}\ dyn\ c\isactrlsub i\ c\isactrlsub o\ h\isactrlsub l\ h\isactrlsub h\ {\isasymtau}{\isacharparenright}\ INV\ I\ h\isactrlsub l\ h\isactrlsub h{\isachardoublequoteclose}\isanewline
\ \ \ \ \isacommand{apply}\isamarkupfalse%
{\isacharparenleft}simp\ only{\isacharcolon}\ mult{\isachardot}assoc{\isacharcomma}\ refinement{\isacharsemicolon}\ {\isacharparenleft}simp{\isacharparenright}{\isacharquery}{\isacharparenright}\isanewline
\ \ \ \ \ \ \ \ \ \isacommand{prefer}\isamarkupfalse%
\ {\isadigit{4}}\ \isacommand{using}\isamarkupfalse%
\ tank{\isacharunderscore}diff{\isacharunderscore}inv\ assms\ \isacommand{apply}\isamarkupfalse%
\ force{\isacharplus}\isanewline
\ \ \ \ \isacommand{using}\isamarkupfalse%
\ tank{\isacharunderscore}inv{\isacharunderscore}arith{\isadigit{1}}\ tank{\isacharunderscore}inv{\isacharunderscore}arith{\isadigit{2}}\ assms\ \isacommand{by}\isamarkupfalse%
\ rel{\isacharunderscore}auto{\isacharprime}\isanewline
\ \ \isacommand{ultimately}\isamarkupfalse%
\ \isacommand{show}\isamarkupfalse%
\ {\isachardoublequoteopen}\isactrlbold {\isacharbrackleft}I\ h\isactrlsub l\ h\isactrlsub h{\isacharcomma}\ I\ h\isactrlsub l\ h\isactrlsub h\isactrlbold {\isacharbrackright}\ {\isasymge}\ tank{\isacharunderscore}dinv\ c\isactrlsub i\ c\isactrlsub o\ h\isactrlsub l\ h\isactrlsub h\ {\isasymtau}{\isachardoublequoteclose}\isanewline
\ \ \ \ \isacommand{by}\isamarkupfalse%
\ auto\isanewline
\isacommand{qed}\isamarkupfalse%
\isanewline
\end{isabellebody}

\noindent The proof incrementally refines the specification of
\isa{tank{\isacharunderscore}dinv} using the laws of $\dR$. As in
Example~\ref{ex:therm-rsol}, after refining the first two assignments,
tactic \isa{refinement} completes the construction of
\isa{ctrl}. Then, the invariant is supplied via lemma
\isa{tank{\isacharunderscore}diff{\isacharunderscore}inv} from
Example~\ref{ex:tank-inv} to construct \isa{dyn}. The final program is then
obtained by transitivity of $\leq$. A more detailed derivation
is also possible. \qed
\end{example}

%%%%%%%%%%%%%%%%%%%%%%%%%%%%%%%%%%%%%%%%%%%%%%%%%%

\section{Evolution Commands for Flows}\label{sec:from-flows}

Finally, we present variants of $\dH$ and $\dR$ that start directly
from flows $\flow:T\to S\to S$ instead of vector fields.  This avoids
checking the conditions of the Picard-Lindel\"of theorem and
simplifies verification proofs considerably.  Instead of
$x'=f\, \&\, G$, we now use the command
$\mathbf{evol}\, \flow\, G$ in hybrid programs and define
\begin{equation*}
  (\mathbf{evol}\, \flow\, G) = \lambda s.\ \gamma^{\flow_s}_G\, s
\end{equation*}
with respect to the guarded orbit of $\flow_s$ along $T$ in $s$. It
then remains to derive a Hoare-style axiom and a refinement law
for such evolution commands. 
\begin{lemma}\label{P:hr-evlfl}
  Let $\flow:T\to S\to S$, where $S$ is a set and $T$ a
  preorder. Then, for $G,P,Q:S\to \bools$,
\begin{gather*}
\{\lambda s\in S.\forall t\in T.\ (\forall
\tau\in {\downarrow}t.\ G\, (\flow_s\, \tau)) \rightarrow P\,
(\flow_s\, t)\}\, \mathbf{evol}\, \flow\, G\, \{P\}, \label{eq:h-evlfl}\tag{h-evlfl}\\
\mathbf{evol}\, \flow\, G \le [\lambda s.\forall t\in T.\ (\forall
\tau\in {\downarrow}t.\ G\, (\flow_s\, \tau))\to Q\, (\flow_s\, t),Q],\label{eq:r-evlf}\tag{r-evlf}\\
(\mathbf{evol}\, \flow\, G) \cdot \left[Q,Q\right] \le [\lambda s. \forall t\in T.\ (\forall
\tau\in {\downarrow}t.\ G\, (\flow_s\, \tau))\to Q\, (\flow_s\, t),Q],\label{eq:r-evlfl}\tag{r-evlfl}\\
\left[Q,\lambda s. \forall t\in T.\ (\forall
\tau\in {\downarrow}t.\ G\, (\flow_s\, \tau))\to Q\, (\flow_s\, t)\right]\cdot (\mathbf{evol}\, \flow\, G) \le [Q,Q].\label{eq:r-evlfr}\tag{r-evlfr}\\
\end{gather*}
\end{lemma}

\begin{example}[Bouncing ball via Hoare logic and refinement]\label{ex:ball-hoare}
  A ball of mass $m$ falls down from height $h\geq 0$, with
  $x$ denoting its position, $v$ its velocity and $g$ its
  acceleration. Its kinematics is modelled by the flow
\begin{isabellebody}
\isanewline
\isacommand{abbreviation}\ %\isamarkupfalse%
%\ ball{\isacharunderscore}flow\ {\isacharcolon}{\isacharcolon}\ {\isachardoublequoteopen}real\ {\isasymRightarrow}\ real\ {\isasymRightarrow}\ {\isacharparenleft}real{\isacharcircum}{\isadigit{2}}{\isacharparenright}\ usubst{\isachardoublequoteclose}\ {\isacharparenleft}{\isachardoublequoteopen}{\isasymphi}{\isachardoublequoteclose}{\isacharparenright}\ \isanewline
%\ \ \isakeyword{where}\ 
{\isachardoublequoteopen}{\isasymphi}\ g\ {\isasymtau}\ {\isasymequiv}\ {\isacharbrackleft}x\ {\isasymmapsto}\ g\ {\isasymcdot}\ {\isasymtau}\ {\isacharcircum}\ {\isadigit{2}}{\isacharslash}{\isadigit{2}}\ {\isacharplus}\ v\ {\isasymcdot}\ {\isasymtau}\ {\isacharplus}\ x{\isacharcomma}\ \ v\ {\isasymmapsto}\ g\ {\isasymcdot}\ {\isasymtau}\ {\isacharplus}\ v{\isacharbrackright}{\isachardoublequoteclose}\isanewline
\end{isabellebody}

% function
%  $\flow:\reals\to\reals^2\to\reals^2$ with
% \begin{equation*}
%    \flow\, t\,
%    \begin{pmatrix}
%      s_x\\
%      s_v
%    \end{pmatrix}
%=
%\begin{pmatrix}
%  s_x+s_v t-\frac{1}{2}gt^2\\
% s_v-g t
%\end{pmatrix},
%\end{equation*}
%where we abbreviate
%$s_x = s\, x$ and $s_v = s\, v$.  
\noindent The ball bounces back elastically
from the ground. This is modelled by a
discrete control that checks for $x=0$ and then flips the
velocity.  Guard $G=(x\geq 0)$ excludes any motion
below the ground. This is modelled by the hybrid
program~\cite{Platzer18}
\begin{isabellebody}
\isanewline
\isacommand{abbreviation}\isamarkupfalse%
\ {\isachardoublequoteopen}bb{\isacharunderscore}evol\ g\ h\ T\ {\isasymequiv}\ \isanewline
\ \ LOOP\
{\isacharparenleft}EVOL\
{\isacharparenleft}{\isasymphi}\ g{\isacharparenright}\
{\isacharparenleft}x\ {\isasymge}\ {\isadigit{0}}{\isacharparenright}\
T{\isacharsemicolon} {\isacharparenleft}IF\ {\isacharparenleft}v\ {\isacharequal}\ {\isadigit{0}}{\isacharparenright}\ THEN\ {\isacharparenleft}v\ {\isacharcolon}{\isacharcolon}{\isacharequal}\ {\isacharminus}v{\isacharparenright}\ ELSE\ skip{\isacharparenright}{\isacharparenright}\ \isanewline
\ \   INV\ {\isacharparenleft}{\isadigit{0}}\ {\isasymle}\ x\ {\isasymand}\ {\isadigit{2}}\ {\isasymcdot}\ g\ {\isasymcdot}\ x\ {\isacharequal}\ {\isadigit{2}}\ {\isasymcdot}\ g\ {\isasymcdot}\ h\ {\isacharplus}\ v\ {\isasymcdot}\ v{\isacharparenright}{\isachardoublequoteclose}\isanewline
\end{isabellebody}
\noindent Its loop invariant conjoins the guard $G$ with a
variant of energy conservation. The correctness
specification and proof with $\dH$ and $\dR$ are then straightforward.
%\begin{align*}
%	\mathsf{Cntrl} &= \IF {(\lambda\, s.\ s_x=0)} {v:=(\lambda\, s.\ - s_v)} \mathit{skip},\\
%	\mathsf{Ball} &= ({\mathbf{evol}\, \flow\, G}\,  {;}\, \mathsf{Cntrl})^\ast.
%\end{align*}
%Its correctness specification is $\{P\}\, \mathsf{Ball}\, \{Q\}$ for $P= (\lambda s.\ s_x = h\land s_v = 0)$ and $Q  = (\lambda s.\ 0\leq s_x\leq h)$; its loop invariant
%\begin{equation*}
%      I = \left(\lambda s.\ 0\le s_x \land \frac{1}{2}s_v^2= g(h - s_x)\right).
%\end{equation*}
%We use Isabelle's \isa{{\isadigit{1}}{\isacharcomma} {\isadigit{2}}{\isacharcolon}{\isacharcolon}{\isadigit{2}}} for $x$ and $v$ respectively. The bouncing ball specification $\{P\}\, \mathsf{Ball}\, \{Q\}$ and its refinement version $\mathsf{Ball}\leq [P,Q]$ are then formalised as below. 
\begin{isabellebody}
\isanewline
\isacommand{lemma}\isamarkupfalse%
\ bouncing{\isacharunderscore}ball{\isacharunderscore}dyn{\isacharcolon}\ \isanewline
\ \ \isakeyword{assumes}\ {\isachardoublequoteopen}g\ {\isacharless}\ {\isadigit{0}}{\isachardoublequoteclose}\ \isakeyword{and}\ {\isachardoublequoteopen}h\ {\isasymge}\ {\isadigit{0}}{\isachardoublequoteclose}\isanewline
\ \ \isakeyword{shows}\ {\isachardoublequoteopen}\ \isactrlbold {\isacharbraceleft}x\ {\isacharequal}\ h\ {\isasymand}\ v\ {\isacharequal}\ {\isadigit{0}}\isactrlbold {\isacharbraceright}\ bb{\isacharunderscore}evol\ g\ h\ T\ \isactrlbold {\isacharbraceleft}{\isadigit{0}}\ {\isasymle}\ x\ {\isasymand}\ x\ {\isasymle}\ h\isactrlbold {\isacharbraceright}{\isachardoublequoteclose}\isanewline
\ \ \isacommand{apply}\isamarkupfalse%
{\isacharparenleft}hyb{\isacharunderscore}hoare\ {\isachardoublequoteopen}{\isactrlU}{\isacharparenleft}{\isadigit{0}}\ {\isasymle}\ x\ {\isasymand}\ {\isadigit{2}}\ {\isasymcdot}\ g\ {\isasymcdot}\ x\ {\isacharequal}\ {\isadigit{2}}\ {\isasymcdot}\ g\ {\isasymcdot}\ h\ {\isacharplus}\ v\ {\isasymcdot}\ v{\isacharparenright}{\isachardoublequoteclose}{\isacharparenright}\isanewline
\ \ \isacommand{using}\isamarkupfalse%
\ assms\ \isacommand{by}\isamarkupfalse%
\ {\isacharparenleft}rel{\isacharunderscore}auto{\isacharprime}\ simp{\isacharcolon}\ bb{\isacharunderscore}real{\isacharunderscore}arith{\isacharparenright}\isanewline
\isanewline
\isacommand{lemma}\isamarkupfalse\ R{\isacharunderscore}bouncing{\isacharunderscore}ball{\isacharunderscore}dyn{\isacharcolon}\isanewline
\ \ \isakeyword{assumes}\ {\isachardoublequoteopen}g\ {\isacharless}\ {\isadigit{0}}{\isachardoublequoteclose}\ \isakeyword{and}\ {\isachardoublequoteopen}h\ {\isasymge}\ {\isadigit{0}}{\isachardoublequoteclose}\isanewline
\ \ \isakeyword{shows}\ {\isachardoublequoteopen}\isactrlbold {\isacharbrackleft}x\ {\isacharequal}\ h\ {\isasymand}\ v\ {\isacharequal}\ {\isadigit{0}}{\isacharcomma}\ {\isadigit{0}}\ {\isasymle}\ x\ {\isasymand}\ x\ {\isasymle}\ h\isactrlbold {\isacharbrackright}\ {\isasymge}\ bb{\isacharunderscore}evol\ g\ h\ T{\isachardoublequoteclose}\isanewline
\ \ \isacommand{apply}\isamarkupfalse%
{\isacharparenleft}refinement{\isacharsemicolon}\ {\isacharparenleft}rule\ R{\isacharunderscore}bb{\isacharunderscore}assign{\isacharbrackleft}OF\ assms{\isacharbrackright}{\isacharparenright}{\isacharquery}{\isacharparenright}\isanewline
\ \ \isacommand{using}\isamarkupfalse%
\ assms\ \isacommand{by}\isamarkupfalse%
\ {\isacharparenleft}rel{\isacharunderscore}auto{\isacharprime}\ simp{\isacharcolon}\ bb{\isacharunderscore}real{\isacharunderscore}arith{\isacharparenright}\isanewline
\end{isabellebody}
\noindent % After applying tactic \isa{hyb{\isacharunderscore}hoare}, the
% remaining proof obligations in the Hoare triple specification can be
% discharged automatically with the arithmetical facts
% (\isa{bb{\isacharunderscore}real{\isacharunderscore}arith}). 
In the
refinement proof, the tactic leaves only the refinement for the
assignment \isa{v\ {\isacharcolon}{\isacharcolon}{\isacharequal}\
  {\isacharminus}v}. This is supplied via lemma
\isa{R{\isacharunderscore}bb{\isacharunderscore}assign} and the
remaining obligations are discharged with the same arithmetical
facts. \qed
\end{example}

%%%%%%%%%%%%%%%%%%%%%%%%%%%%%%%%%%%%%%%%%%%%%%%%%%

\section{Conclusion}\label{sec:conclusion}

We have contributed new methods and Isabelle components to an open
modular semantic framework for verifying hybrid systems that so far
focussed on predicate transformer semantics~\cite{MuniveS19}; more
specifically the first standalone Hoare logic $\dH$ for hybrid
programs, the first Morgan-style refinement calculus $\dR$ for such
programs, more generic state spaces modelled by lenses, improved
Isabelle syntax for correctness specifications and hybrid programs,
and increased proof automation via tactics.  These components support
three workflows based on certifying solutions to Lipschitz-continuous
vector fields, reasoning with invariant sets for continuous vector
fields, and working directly with flows without certification.

Compared to the standard $\dL$ toolchain, $\dH$ and $\dR$ are
simple. They emphasise the natural mathematical style of reasoning
about dynamical systems, with minimal conceptual overhead relative to
standard Hoare logics and refinement calculi. $\dH$, in particular,
remains invisible and is only used for automated verification
condition generation. The modular approach with algebras and a shallow
embedding has simplified the construction of these verification
components and made it incremental relative to extant ones. Our
framework is not only open to use any proof method and mathematical
approach supported by Isabelle, it should also allow adding new
methods, for instance based on discrete dynamical systems, hybrid
automata or duration calculi~\cite{LiuLQZZZZ10}, or integrate CAS's
for finding solutions. It should be equally straightforward to
formalise $\dH$ and $\dR$ based on other Hoare logics in Isabelle with
our hybrid store models.

The relevance of $\dH$ and $\dR$ to hybrid systems verification is
further evidenced by the fact that such approaches are not new: A
hybrid Hoare logic has been proposed by Liu et al.~\cite{LiuLQZZZZ10}
for a duration calculus based on hybrid CSP and been widely used
since. It is conceptually very different from $\dH$ and $\dL$. A
differential refinement logic based on $\dL$ has been developed as
part of Loos' PhD work~\cite{LoosP16}.  It uses a proof system with
inference rules for reasoning about inequalities between $\KAT$
expressions, which are interpreted in a rather non-standard way as
refinements between hybrid programs. It differs substantially from the
calculi developed by Back and von Wright~\cite{BackW98},
Morgan~\cite{Morgan94} and others, and thus from the predicate
transformer algebras in~\cite{MuniveS19} and from $\dR$.  The relative
merits of these approaches remain to be explored.

The expressivity and complexity gap between Hoare logic and predicate
transformer semantics is particularly apparent within algebra. The
weakest liberal precondition operator cannot be expressed in
$\KAT$~\cite{Struth18}.  The equational theory of $\KAT$, which
captures propositional Hoare logic, is PSPACE
complete~\cite{KozenCS96}, that of modal Kleene algebra, which yields
predicate transformers, is in EXPTIME~\cite{MollerS06}.

Finally, while $\KAT$ and $\rKAT$ are convenient starting points for
building program construction and verification components for hybrid
programs, the simple and more general setting of Hoare
semigroups~\cite{Struth18} would support developing hybrid Hoare
logics for total program correctness---where balls may bounce
forever---or even for multirelational
semantics~\cite{FurusawaS16,FurusawaS15} as needed for differential
game logic~\cite{Platzer18}. This, however, is left for future work.

\paragraph{Acknowledgements.}
Second author is sponsored by CONACYT's scholarship no. 440404.

%%%%%%%%%%%%%%%%%%%%%%%%%%%%%%%%%%%%%%%%%%%%%%%%%%

\bibliographystyle{abbrv}
\bibliography{ms}

\end{document}